\let\kernelLabel\label
\documentclass[aps,prx,reprint,superscriptaddress,longbibliography]{revtex4-2}
\makeatletter
\@ifl@t@r\fmtversion{2023-06-01}{\let\label\kernelLabel}{}
\makeatother

\usepackage[utf8]{inputenc}
\usepackage[T1]{fontenc}
\usepackage{lmodern}
\usepackage{amsmath,amssymb,amsthm}
\usepackage{booktabs}
\usepackage{graphicx}
\usepackage{microtype}
\microtypesetup{stretch=30}
\usepackage{etoolbox}
\makeatletter
\patchcmd{\@bibdataout@aps}{author="08"}{author="48"}{}{\errmessage{Cannot update APS bibliography author control}}
\patchcmd{\@bibdataout@aps}{author="08"}{author="48"}{}{\errmessage{Cannot update APS bibliography author control}}
\makeatother
\usepackage{xurl}
\usepackage[colorlinks=true,linkcolor=blue,citecolor=blue,urlcolor=blue]{hyperref}
\newcommand{\DatasetDOIDisplay}[1]{, DOI: \href{https://doi.org/#1}{\nolinkurl{#1}}}
\hypersetup{pdftitle={Sufficient quantum provenance: retained fields and certified recording precision},pdfauthor={George Kourousias},pdfkeywords={Benchmarking, Hellinger distance, metadata, provenance, quantum computing, quantum software, recording precision, reproducibility}}

\newtheorem{definition}{Definition}
\newtheorem{proposition}{Proposition}
\newcommand{\FU}{\mathcal{F}}
\newcommand{\eps}{\varepsilon}
\newcommand{\risk}{\mathcal{R}}
\newcommand{\dtv}{d_{\mathrm{TV}}}

\begin{document}
\title{Sufficient quantum provenance: retained fields and certified recording precision}
\author{George Kourousias}
\email{george.kourousias@elettra.eu}
\affiliation{Elettra Sincrotrone Trieste, Scientific \& Quantum Computing Unit, Strada Statale 14--km 163.5, 34149 Trieste, Italy}
\affiliation{University of Trieste, Image Processing Laboratory, Department of Engineering and Architecture, Piazzale Europa 1, 34127 Trieste, Italy}
\affiliation{National Centre of Scientific Research ``Demokritos'', Institute of Quantum Computing and Quantum Technology, 15341 Agia Paraskevi, Athens, Greece}
\author{Sergio Carrato}
\affiliation{University of Trieste, Image Processing Laboratory, Department of Engineering and Architecture, Piazzale Europa 1, 34127 Trieste, Italy}
\author{Roberto Pugliese}
\affiliation{Elettra Sincrotrone Trieste, Scientific \& Quantum Computing Unit, Strada Statale 14--km 163.5, 34149 Trieste, Italy}
\author{Aljo\v{s}a Hafner}
\affiliation{Elettra Sincrotrone Trieste, Scientific \& Quantum Computing Unit, Strada Statale 14--km 163.5, 34149 Trieste, Italy}
\author{Francesco Guzzi}
\affiliation{Elettra Sincrotrone Trieste, Scientific \& Quantum Computing Unit, Strada Statale 14--km 163.5, 34149 Trieste, Italy}
\author{Matteo Bill\`e}
\affiliation{Elettra Sincrotrone Trieste, Scientific \& Quantum Computing Unit, Strada Statale 14--km 163.5, 34149 Trieste, Italy}
\author{Fulvio Bill\`e}
\affiliation{Elettra Sincrotrone Trieste, Scientific \& Quantum Computing Unit, Strada Statale 14--km 163.5, 34149 Trieste, Italy}
\author{Panagiotis Dimitrakis}
\affiliation{National Centre of Scientific Research ``Demokritos'', Institute of Quantum Computing and Quantum Technology, 15341 Agia Paraskevi, Athens, Greece}
\date{25 November 2026}
\begin{abstract}
Quantum executions that share a provenance record can still produce different output distributions. We determine which execution details a record must retain, and at what precision, to guarantee agreement within a declared Hellinger tolerance. Under a specified channel model, a certificate combining independent stochastic mechanisms and coherent rotation differences fixes the recording precision without output simulation or explicit dependence on register dimension. For four-qubit quantum approximate optimization at depth one, a 13-bit comparison record suffices at tolerance 0.05 over a declared continuous noise range, uniformly in programmed angles for each fixed task, against 19 bits for a gate-by-gate sum over the same bin family. Pairwise separation constraints characterize sufficient field sets on finite context classes and expose omissions that pass individually but fail jointly. Separately, six-qubit measurements on an IBM Heron processor show, over six tested contexts, that raw outputs need a two-label record, supplied by qubit layout, whereas outputs after a fixed application decoder need no additional context bits, at tolerance 0.15 with 97\% simultaneous confidence under stationary, independent shots. Execution records can thereby be designed to carry checkable agreement guarantees.
\end{abstract}
\maketitle

\section{From preservation to a guarantee}
Physical implementations of the same quantum task can produce different outcome laws. Which execution details must be retained, and at what resolution, for every admissible pair sharing a record to agree within a declared tolerance? Preserved programs, data and environments make computational research inspectable~\cite{Peng2011}, but do not themselves specify this quantitative agreement promise. An empirical study of quantum-computing research documents recurring gaps in artefact availability, environment specification and executability~\cite{Koster2026}. We address a complementary requirement: the comparison guarantee supported by an execution record once the task and admissible contexts are specified.

Our main contribution constructs sufficient recording-bin widths from a specified quantum-channel model. Within a common circuit structure, the certificate combines independent stochastic channel mixtures and coherent rotation differences to bound every admissible pair assigned the same record, without simulating the measured outcome distributions. For the four-qubit quantum approximate optimization algorithm (QAOA) at depth one, 13 comparison bits suffice at Hellinger tolerance $0.05$ over a declared continuous noise box. The same bound holds for every common choice of programmed angles, with task identity fixed within each comparison. It has no explicit dependence on register dimension.

A finite QAOA-1 example makes the need for joint record design concrete. Omitting routing alone or angle bias alone gives computed worst-case Hellinger discrepancies of $0.0417$ and $0.0497$, respectively. Each passes a separate check against the full tolerance $0.05$, but omitting both while retaining layout, gate noise and readout gives $0.0514$. Separate full-tolerance omission checks therefore do not certify their joint omission. The record criterion instead controls all pairs assigned the same record. The full values, witness and a conservative composition bound appear in Section~\ref{sec:study} and Supplementary Section~S1.4.

Dasgupta and Humble connect device characterization to quantum-circuit reproducibility and stability, bounding noisy-versus-ideal output discrepancies~\cite{Dasgupta2022} and changes in noise-averaged observables induced by time-varying distributions of noise parameters~\cite{Dasgupta2024}. Their results already translate outcome tolerances into constraints on characterized noise. Here, the design object is the retained representation of an execution class, constrained by its largest within-record output-law discrepancy. This yields retention constraints for discrete compilation fields such as layout and routing as well as bin widths for continuous parameters. Rudinger et al. detect and quantify context dependence in circuit outcomes~\cite{Rudinger2019}; our framework constructs retained records whose hidden discrepancies meet a declared tolerance. Accreditation addresses accuracy relative to an ideal target~\cite{Mills2026}, whereas the present guarantee concerns agreement between admissible implementations.

The surrounding provenance infrastructure includes PROV~\cite{Moreau2013}, research-object packaging~\cite{SoilandReyes2022}, quantum-context capture and workflows~\cite{Weder2021,Weder2023}, structured circuit records~\cite{Huenchuleo2026}, and Qiskit compiler tracking~\cite{Stirbu2026}. Mathematical connections include predictive compression~\cite{Tishby1999}, discernibility-based attribute reduction~\cite{Skowron1992}, functional compression~\cite{Doshi2010}, and Fisher information as a measurement-resolution constraint~\cite{Braunstein1994,Paris2009}. 

This paper makes three contributions.
\begin{enumerate}
\item A joint stochastic--coherent channel certificate (Proposition~\ref{prop:channel}) that sets sufficient recording precision for continuous parameters without output simulation. For QAOA-1 it yields a 13-bit record where a gate-by-gate sum yields 19 bits.
\item A separation certificate (Proposition~\ref{prop:separation}) that characterizes sufficient field sets on finite context classes. It exposes omissions that pass individually but fail jointly, and exhaustive enumeration certifies finite-family minima.
\item A hardware confirmation on an IBM Heron processor in which a fixed application decoder lowers the certified minimum label count from two to one at a common tolerance.
\end{enumerate}
 Detailed proofs and supporting methods are provided in the Supplementary Information; the dataset supplies the complete numerical evidence and verification code~\cite{QuantumProvenanceDataset}.
\section{What a sufficient record preserves}
\label{sec:criterion}

Fix a logical task, its parameters, and a common measurement alphabet $\mathcal{B}$. Let $\FU$ be a finite set of candidate context fields and $\Omega\subseteq\prod_{k\in\FU}\mathrm{Vals}(k)$ a declared, non-empty set of admissible complete contexts. Dependencies between fields are allowed: $\Omega$ need not be a Cartesian product. Each $z\in\Omega$ specifies an outcome law $p_z$ on $\mathcal{B}$: the probability distribution over a single measurement outcome under that complete context. Repeated runs governed by the same law need not return the same bitstring; a finite-shot histogram is an estimate of the law, not the comparison object used in Sections~\ref{sec:criterion}--\ref{sec:study}. In the simulation study $p_z$ is the raw, single-shot bitstring law of a specified quantum channel, not a mitigation quasi-distribution. Section~\ref{sec:scope} treats finite-shot uncertainty, and Section~\ref{sec:hardware} compares raw and deterministically decoded laws from hardware.

Retaining fields $S\subseteq\FU$ writes the projection $r_S(z)=z|_S$. The record groups contexts that it cannot distinguish. For the Hellinger distance
\begin{equation}
 \begin{aligned}
 H(p,q)&=\frac{1}{\sqrt{2}}\left\|\sqrt{p}-\sqrt{q}\right\|_2,\\
 \risk_\Omega(S)&=\sup_{\substack{z,z'\in\Omega\\r_S(z)=r_S(z')}}H(p_z,p_{z'}),
 \end{aligned}
 \label{eq:risk}
\end{equation}
the second quantity is the largest diameter of any record cell: the largest outcome-law difference that can remain hidden behind an identical retained record. We call this quantity the risk of $S$: the worst discrepancy the record can hide. Record equality defines the partition; closeness within tolerance need not be transitive. Equivalently, $\risk_\Omega(S)$ asks how differently two executions could behave while carrying the same retained context.

The separation construction works with any declared metric or application-specific pseudometric. We use Hellinger distance because it supports the metric and processing properties used below, has been used previously to quantify quantum-circuit reproducibility~\cite{Dasgupta2022,Dasgupta2024}, and, as Section~\ref{sec:precision} shows, is the distance whose local geometry is the Fisher information of the outcome law.

\begin{definition}[Sufficient retention]
A field set $S$ is $\eps$-sufficient on $\Omega$ when $\risk_\Omega(S)\leq\eps$. It is inclusion-minimal if none of its proper subsets is sufficient. For declared positive field costs $c_k$, a minimum-cost sufficient set solves
\begin{equation}
 C_\Omega(\eps)=\min_{S\subseteq\FU}
 \left\{\sum_{k\in S}c_k:\risk_\Omega(S)\leq\eps\right\}.
 \label{eq:cost}
\end{equation}
\end{definition}

When several sets attain the minimum cost, we select the one with the smallest risk, and break any remaining tie by a fixed enumeration order. The margin $\eps-\risk_\Omega(S)$ has a direct use: Section~\ref{sec:scope} shows that a policy remains sufficient under any perturbation of every law by at most half of it, and Section~\ref{sec:study} shows that the choice changes the outcome of the archived parameter checks.

For scale, $\eps=0.05$ implies $\dtv(p_z,p_{z'})\leq\sqrt{2}\eps\approx0.071$ for matching records. The comparison can instead concern a future transcript of $N$ independent shots. Multiplicativity of the affinity $\mathcal A(p,q)=1-H^2(p,q)$ gives the exact equivalence
\begin{equation}
 \begin{gathered}
 \risk^{(N)}(r)=\sqrt{1-(1-\risk(r)^2)^N},\\
 \risk^{(N)}(r)\leq\tau\ \Longleftrightarrow\
 \risk(r)\leq\eps_N(\tau),\\
 \eps_N(\tau):=\sqrt{1-(1-\tau^2)^{1/N}}.
 \end{gathered}
 \label{eq:transcriptmain}
\end{equation}
Here $r$ may be a field projection or any deterministic record; the identity also holds for full count histograms. At fixed $0<\tau<1$, $\eps_N(\tau)\sim\sqrt{-\log(1-\tau^2)/N}$; the further approximation $\eps_N(\tau)\approx\tau/\sqrt N$ additionally requires small $\tau$. The future comparison length $N$ differs from the sample size used to estimate a law.

The guarantee is operational: it concerns the outcome laws of executions that share a record. Minimum-cost solutions are inclusion-minimal, but the converse need not hold; either can be non-unique. Recording more fields subdivides cells, so $S\subseteq T$ implies $\risk_\Omega(T)\leq\risk_\Omega(S)$. Enlarging the admissible class can only increase the risk of a fixed policy. These two monotonicities put the qualification where it belongs: a sufficient record remains sufficient when it remembers more, not necessarily when it is asked to cover more. Matching records guarantee agreement within tolerance; non-matching records do not establish disagreement.

The optimization concerns only the context fields used to decide whether two executions are comparable. The task, the model and software definitions, recording dictionaries, the generated circuit and the outcome data are retained separately as the evidential archive, and every bit count in this paper refers to the comparison record alone. Each complete context generates its own circuit and law.

\section{Continuous precision from quantum channels}
\label{sec:precision}

A continuously valued field cannot be retained exactly in a finite digital record. Let $r(z)$ instead contain declared bin labels and any retained discrete fields, and define $\risk(r)$ by replacing $r_S$ with $r$ in Eq.~\eqref{eq:risk}. Bin labels define equality; their physical widths must support the promised outcome-law diameter. The following certificate uses the channel description, rather than a sampled sensitivity envelope.

Consider two executions with the same input state, circuit structure, and final measurement. A stochastic location $g$ applies one of two fixed completely positive trace-preserving maps, with branch probability $p_g$ in one execution and $p'_g$ in the other. Flags at different locations are independent. Between these locations the executions can differ in Pauli rotations $R_{\nu_j}(\alpha_j)=\exp(-i\alpha_j\sigma_{\nu_j}/2)$, including rotations on subsystems of a larger register. Put
\begin{equation}
 \begin{aligned}
 a_{\rm st}&=\prod_g\!\left[\sqrt{p_gp'_g}+\sqrt{(1-p_g)(1-p'_g)}\right],\\
 D&=\frac12\sum_j|\alpha_j-\alpha'_j|.
 \end{aligned}
 \label{eq:resources}
\end{equation}
The mixture branches, rotation axes, and all other channels are common to the two executions; an unchanged noisy gate can be included among those common channels.

\begin{proposition}[Joint stochastic--coherent certificate]
Under these assumptions, the measured outcome laws obey
\begin{equation}
 H^2(p_z,p_{z'})\leq 1-a_{\rm st}\cos\!\left(\min\{D,\pi/2\}\right).
 \label{eq:joint}
\end{equation}
The bound is independent of register dimension and of the common programmed angles. It is attained by a family with observable noise flags and commuting rotations, for displacements within a monotone measurement sector.
\label{prop:channel}
\end{proposition}

To prove the bound, retain the complete latent flag string $f$ temporarily, with laws $P_f,Q_f$ and conditional final states $\rho_f,\sigma_f$. For root fidelity $F(\rho,\sigma)=\|\sqrt\rho\sqrt\sigma\|_1$, discarding flags and then measuring gives
\begin{equation}
 \mathcal A(p_z,p_{z'})\geq\sum_f\sqrt{P_fQ_f}\,F(\rho_f,\sigma_f).
 \label{eq:flagproof}
\end{equation}
Along a fixed branch string, contractivity and the triangle inequality for the Bures angle bound the accumulated angle by $D$: one rotation contributes at most $|\alpha_j-\alpha'_j|/2$ in the nontrivial regime. Thus every conditional fidelity is at least $\cos(\min\{D,\pi/2\})$. The affinity of the independent flags factorizes into $a_{\rm st}$, proving Eq.~\eqref{eq:joint}. The proof uses standard fidelity and classical-simulation tools~\cite{Watrous2018,Demkowicz2012,Yuan2017}; Supplementary Section~S1.10 gives it in full, with a general conditional version and the attaining construction. All budgets derived from Eq.~\eqref{eq:joint} are sufficient: they give upper bounds on the required comparison-record length.

Writing $\theta_g=\arcsin\sqrt{p_g}$ makes $a_{\rm st}=\prod_g\cos(\theta_g-\theta'_g)$. For small displacements,
\begin{equation}
 \begin{split}
 H^2(p_z,p_{z'})\leq{}&\frac12\sum_g(\Delta\theta_g)^2\\
 &+\frac18\left(\sum_j|\Delta\alpha_j|\right)^2+O(\|\Delta\|^4).
 \end{split}
 \label{eq:localchannel}
\end{equation}
Independent stochastic contributions combine quadratically, whereas a shared coherent bias can accumulate across rotations before squaring. Crucially, stochastic and coherent contributions also compose in the squared bound. Summing a separate Hellinger allowance for every noisy gate loses this structure. This joint structure, familiar from quantum metrology, is what reduces the QAOA-1 record below from 19 bits to 13.

\begin{figure*}[!t]
\centering
\includegraphics[width=\textwidth]{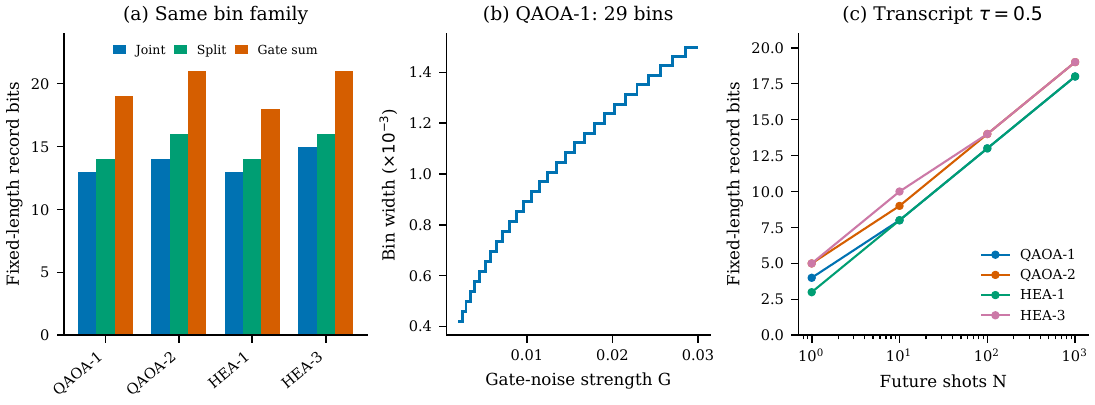}
\caption{Constructive recording budgets from channel structure. (a) Fixed-length record bits at single-shot tolerance $0.05$ for the joint certificate, separate stochastic/coherent composition, and a gatewise sum, with all three using the same bin family. (b) QAOA-1 gate-noise bin widths increase with $G$ under equal intervals in $\arcsin\sqrt{1.25G}$. (c) Precision budgets for an $N$-shot transcript tolerance $\tau=0.5$, using the exact $\eps_N(\tau)$. Lines connect calculated allocations.}
\label{fig:channels}
\end{figure*}

\subsection{A uniform certificate for the study circuits}
We illustrate the certificate with four-qubit QAOA and hardware-efficient ansatz circuits compiled onto a four-site line (Section~\ref{sec:study}). Three initial layouts and two routing rules give six compiled structures per circuit family and depth; $K$ denotes an upper bound on their physical CNOT counts, including routing overhead. Each physical CNOT carries a stochastic channel-mixture probability $w_gG$, with $w_g\leq w_*=1.25$; measurement includes four independent bit flips of probability $M$, and $m$ rotations share an additive angle bias $A$. Retain layout and routing, and partition
\[
 \begin{gathered}
 G\in[0.002,0.030],\qquad M\in[0.005,0.060],\\
 A\in[0,0.08]
 \end{gathered}
\]
into $K_G,K_M,K_A$ bins of equal full width in $x=\arcsin\sqrt{w_*G}$, $y=\arcsin\sqrt M$, and $A$, respectively. Let $X,Y,W_A$ be the three full range widths and put $D_A=\min\{mW_A/(2K_A),\pi/2\}$. Since $|\Delta\arcsin\sqrt{w_gG}|\leq|\Delta x|$, Eq.~\eqref{eq:joint} gives the whole-cell bound
\begin{equation}
 \begin{gathered}
 \risk(r)\leq B(K_G,K_M,K_A),\\
 B^2:=1-\cos^{K}(X/K_G)\cos^4(Y/K_M)\cos D_A.
 \end{gathered}
 \label{eq:boxbound}
\end{equation}
The sensitivity inequality for $w_g\leq w_*$ is proved in the Supplement. No Fisher supremum, grid coverage assumption, or outcome simulation is used. Because layout and routing are retained, structure-dependent binning can replace $K$ by the corresponding compiled CNOT count; we use a common allocation across structures to simplify the recording dictionary. The record omits the law-invariant sampler label. With the common allocation, its $6K_GK_MK_A$ possible labels require $\lceil\log_2(6K_GK_MK_A)\rceil$ bits in a joint fixed-length encoding of the named labels; separately padded fields can require more bits.

\begin{table*}[!ht]

\centering\small
\begin{tabular}{@{}lrrlrr@{}}
\toprule
Circuit & $K$ & $m$ & $(K_G,K_M,K_A)$ & Bound & Record bits \\
\midrule
QAOA-1 & 56 & 4 & (29, 8, 4) & 0.049715 & 13 \\
QAOA-2 & 112 & 8 & (36, 9, 8) & 0.049748 & 14 \\
HEA-1 & 15 & 8 & (13, 9, 8) & 0.049993 & 13 \\
HEA-3 & 45 & 16 & (24, 9, 15) & 0.049980 & 15 \\
\bottomrule
\end{tabular}

\caption{Analytically sufficient precision over the entire continuous noise box at $\eps=0.05$, with layout and routing retained. Counts are obtained from all six compiled structures. Label allocations are selected by a finite integer search under Eq.~\eqref{eq:boxbound}; their feasibility is checked by outward interval arithmetic. The bound is a conservative channel budget within the specified bin family. Each certificate holds for arbitrary programmed angles of that circuit family, with the logical task held fixed within every comparison.}
\label{tab:channel}
\end{table*}

For QAOA-1, $K=56$ and $m=4$ suffice. The $(29,8,4)$ allocation has $B<0.049716$ and 5,568 record labels, so 13 bits suffice. A gate-by-gate Hellinger sum, optimized over the same bin family, yields a 19-bit construction; separating joint stochastic noise from coherent error by a triangle inequality yields 14 bits. Across the four circuits, the joint construction uses 13--15 bits (Table~\ref{tab:channel}). The bound's uniformity in programmed angles is stronger in scope than the finite seed checks below, but it retains routing and is more conservative about precision. Figure~\ref{fig:channels} summarizes these budgets and illustrates the associated bin widths and transcript-length dependence.

\subsection{Statistical resolution and geometry}
Equation~\eqref{eq:transcriptmain} turns Eq.~\eqref{eq:boxbound} into a transcript certificate by setting $\eps=\eps_N(\tau)$. At $\tau=0.5$, the QAOA-1 construction requires 13 bits for $N=100$ and 18 bits for $N=1000$. For these three continuous fields, the construction has cost $\tfrac32\log_2N+O(1)$ as $N\to\infty$.

Actual measurements can be less sensitive. Along a smooth path of positive outcome laws with Fisher information $I(t)$,
\begin{equation}
 H(p_{z(0)},p_{z(1)})\leq\frac{1}{2\sqrt2}\int_0^1\sqrt{I(t)}\,dt.
 \label{eq:pathbound}
\end{equation}
Differentiating $\sqrt{p_t}$ proves the bound~\cite{Braunstein1994,Paris2009}. Its local width scale is $\sqrt8\,\eps/\sqrt I$, a statistical resolution scale. Unlike a sampled sensitivity envelope, Eq.~\eqref{eq:joint} supplies a uniform guarantee over the continuous box and yields a direction-dependent Fisher upper bound throughout the model domain (Supplementary Section~S1.10).

\subsection{The cost of preserving field names}
For a finite class, arbitrary sufficient encodings are colourings of the graph of tolerance-violating pairs~\cite{Doshi2010}. Matching clique and colouring certificates~\cite{TomitaSeki2003,Brelaz1979} show that seven labels suffice and are necessary for the QAOA-1 design grid at $\eps=0.05$, versus 81 labels for its selected four-field record: three versus seven fixed-length bits. A worked rotation example in Supplementary Section~S1.9 shows that separately encoding $m$ offsets can cost $\tfrac m2\log_2N+O(1)$ bits although encoding their observable sum costs $\tfrac12\log_2N+O(1)$. The separation concerns permitted encodings; related parametric compression rates are established~\cite{Yang2018}. Naming fields preserves mechanism-specific distinctions that a synthesized label need not preserve.

\section{Field retention and separation certificates}
\label{sec:separation}

One natural strategy is to remove a field from a reference record and compare the reference law with the mixture obtained by averaging over that missing field. Such an ablation measures a local loss under a chosen distribution of contexts. It need not measure the diameter that the record conceals.

A single qubit makes the distinction exact. Let two binary context fields $a,b$ select the circuit $R_y(\pi/2+ab\pi/2)|0\rangle$. Three contexts yield $p(1)=1/2$; the fourth yields $p(1)=1$. At reference $(a,b)=(0,0)$, either single-field ablation has zero deficit. Even forgetting both fields and mixing the four contexts uniformly gives only $H=0.0893$ relative to the reference. Nevertheless, the empty record hides a pair at distance $0.5412$. At tolerance $0.1$, both fields are necessary. Averaging has concealed a failure, not certified its absence. Supplementary Section~S1 gives the exact expressions.

Replacing a reference mixture with a worst-case single-field check does not by itself solve the composition problem. Write $s_k=\risk_\Omega(\FU\setminus\{k\})$ and let the \emph{singleton-threshold rule} retain $B_\eps=\{k:s_k>\eps\}$. Each retained field is forced by a violating pair differing in that field alone. But $s_k\leq\eps$ for every omitted field does not imply $\risk_\Omega(B_\eps)\leq\eps$: those checks retain all other fields, whereas the proposed policy omits them jointly. On Cartesian support, a record cell is a box in context space, but its image in the space of outcome laws can be curved. Single-field checks control coordinate changes; the guarantee concerns every pair in the cell. The numerical example below tests this joint omission explicitly; Supplementary Section~S1.4 states the scope of single-field composition bounds.

There is a finite alternative that checks the guarantee itself. For every pair with $H(p_z,p_{z'})>\eps$, define its distinguishing fields
\begin{equation}
 \Delta(z,z')=\{k\in\FU:z_k\ne z'_k\}.
 \label{eq:distinguish}
\end{equation}
\begin{proposition}[Separation certificate]
For finite $\Omega$, $S$ is $\eps$-sufficient if and only if it intersects every $\Delta(z,z')$ belonging to a tolerance-violating pair. Thus Eq.~\eqref{eq:cost} is a weighted hitting-set problem. A sufficient set is inclusion-minimal if and only if every retained field $k$ has a violating pair with $\Delta(z,z')\cap S=\{k\}$.
\label{prop:separation}
\end{proposition}

The proof is the definition read in both directions: an unseparated violating pair inhabits one record cell; separating every such pair excludes all violations. The second statement supplies a concrete answer to why a field is needed: remove it and this pair becomes indistinguishable. It is the distribution-valued counterpart of a discernibility construction~\cite{Skowron1992}. These witnesses certify inclusion-minimality. Global cost optimality requires excluding every cheaper candidate, which the study below does by enumerating all field sets.

\section{An executable finite-class study}
\label{sec:study}

This section compares exhaustive minimum-cost selection with two natural omission rules on four small circuits; both rules can fail the guarantee they are meant to support.

We compute raw outcome laws by density-matrix propagation for two hardware-efficient ansatz circuits, at one and three entangling layers, and two four-cycle MaxCut QAOA circuits, at depths one and two. These are small representatives of established circuit families~\cite{Kandala2017,Farhi2014}. Each design instance has 324 contexts: three initial layouts, two explicit routing rules on a four-site line, three two-qubit depolarizing channel-mixture strengths, three readout-error probabilities, three coherent rotation biases, and two sampler-seed labels. The seed is a negative control: it cannot change the underlying outcome law. Single-qubit depolarization is fixed. The simulator, grid, parameters, observables, and routing rules are fully specified in Supplementary Section~S2.

For each design instance we evaluate all $2^6=64$ field sets, using unit field costs. The primary tolerance is $\eps=0.05$, with an exploratory sweep from $0.01$ to $0.20$. We compare the reference-mixture rule with the singleton-threshold rule on the same laws. All probabilities come from exact channel propagation in floating point, without shot sampling.

\begin{table*}[!ht]

\centering
\small
\begin{tabular}{@{}lcrrrrrr@{}}
\toprule
 & & \multicolumn{2}{c}{Minimum} & \multicolumn{2}{c}{Mixture} & \multicolumn{2}{c}{Singleton} \\
\cmidrule(lr){3-4}\cmidrule(lr){5-6}\cmidrule(l){7-8}
Task & Selected minimum set & Fields & Risk & Fields & Risk & Fields & Risk \\
\midrule
HEA-1 & $L,R,G,M,A$ & 5 & 0.0000 & 1 & \textbf{0.2262} & 5 & 0.0000 \\
HEA-3 & $L,R,G,M,A$ & 5 & 0.0000 & 1 & \textbf{0.2924} & 5 & 0.0000 \\
QAOA-1 & $L,G,M,A$ & 4 & 0.0417 & 1 & \textbf{0.1215} & 3 & \textbf{0.0514} \\
QAOA-2 & $L,G,M,A$ & 4 & 0.0250 & 1 & \textbf{0.1462} & 4 & 0.0250 \\
\bottomrule
\end{tabular}

\caption{Finite-class retention at $\eps=0.05$. Letters denote layout (L), routing (R), gate noise (G), readout (M), and angle bias (A); sampler seed is never necessary. Risk is the maximum record-cell Hellinger diameter; bold values violate the tolerance. Zero means only the law-invariant seed is omitted. Among minimum-cost sets the one with the smallest risk is selected; QAOA-1 has two, and both are reported in the text.}
\label{tab:records}
\end{table*}

\begin{figure*}[!t]
  \centering
  \includegraphics[width=\textwidth]{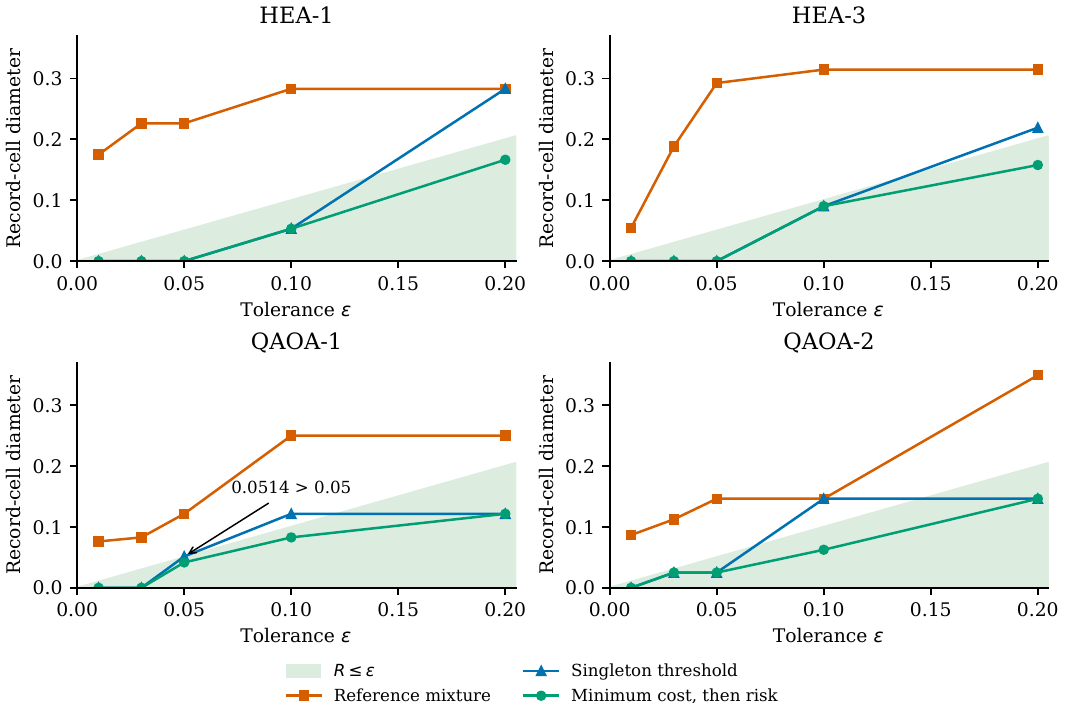}
  \caption{Retention policies evaluated against the guarantee they are intended to support. Each panel shows the achieved record-cell diameter $\risk_\Omega(S)$ for the reference-mixture rule, singleton-threshold rule, and exhaustive minimum-cost sufficient set over the five declared tolerances. The shaded region marks $\risk_\Omega(S)\leq\eps$; points above it violate their nominal tolerance. The annotated QAOA-1 case at $\eps=0.05$ is the compositional failure discussed in the text: the singleton rule attains $0.0514$ although routing and angle bias each pass the one-field omission check separately. Values are from the complete baseline sweep in the electronic dataset; all minimum-cost curves use the residual-risk tie-break. Connecting lines and the shaded boundary are visual guides across the discrete tolerance values.}
  \label{fig:baselines}
\end{figure*}

Table~\ref{tab:records} gives the headline $\eps=0.05$ slice. Figure~\ref{fig:baselines} extends the same comparison across the full declared tolerance sweep. Across this finite grid, the reference-mixture rule violates the intended guarantee throughout, while the singleton-threshold rule is much closer but can still cross the boundary when individually acceptable omissions are combined.

The reference-mixture rule retains only readout for the ansatz instances and only gate noise for QAOA, leaving diameters of $0.1215$--$0.2924$ (Table~\ref{tab:records}). For HEA-1, routing has zero local score at the identity layout but a single-field worst-case distance of $0.1211$ elsewhere. The singleton rule therefore detects a failure that the reference misses.

QAOA-1 exposes the remaining gap. At $\eps=0.05$, singleton violations force L, G, M, but omitting R alone gives $s_{\mathrm R}=0.0416650$ and omitting A alone gives $s_{\mathrm A}=0.0496812$. Omitting both gives $\risk_\Omega(\{\mathrm{L,G,M}\})=0.0514011$. As in an engineering error budget, two omissions can each lie within the full tolerance when assessed separately while their joint omission exceeds it; the distances themselves need not add. The witness has reversed layout, gate-noise strength $0.002$, readout error $0.005$, and seed $7$; it differs only in routing (move control versus move target) and angle bias ($0$ versus $0.08$). It requires R or A in addition to the forced fields. Both resulting four-field sets are sufficient: $\{\mathrm{L,R,G,M}\}$ has risk $0.0496812$, and $\{\mathrm{L,G,M,A}\}$ has risk $0.0416650$. The residual-risk rule selects the second; both remain archived with necessity witnesses. A conservative single-field composition bound that avoids this failure is given in Supplementary Section~S1.4.

The comparison target matters too. For HEA-1, retaining A alone meets a $Z_0$-expectation tolerance of $0.03$ with semantic risk $0.0260182$ rounded upward, yet leaves distributional risk $0.2759$, versus $0.2830$ for empty retention. The two criteria have different targets and units.

\begin{table*}[!ht]

\centering
\small
\begin{tabular}{@{}llrrrr@{}}
\toprule
Task & Policy & Design risk & Allowance & Max.\ risk elsewhere & Passes \\
\midrule
HEA-1 & $L,R,G,M,A$ & 0.0000 & 0.0250 & 0.0000 & 3/3 \\
HEA-3 & $L,R,G,M,A$ & 0.0000 & 0.0250 & 0.0000 & 3/3 \\
QAOA-1 & $L,R,G,M$ & 0.0497 & 0.0002 & \textbf{0.0593} & 9/13 \\
QAOA-1 & $L,G,M,A$ & 0.0417 & 0.0042 & 0.0444 & 13/13 \\
QAOA-2 & $L,G,M,A$ & 0.0250 & 0.0125 & 0.0384 & 13/13 \\
\bottomrule
\end{tabular}

\caption{Every minimum-cardinality policy at $\eps=0.05$ applied, without reselection, to the other archived parameter instances of its workload (seeds 43--45 for all workloads, 46--55 for QAOA). Allowance is $(\eps-\risk_\Omega(S))/2$, the perturbation of every law that the policy provably tolerates. Bold values exceed the tolerance. Each distance compares contexts within one fixed task; no distance compares different task parameters.}
\label{tab:margin}
\end{table*}

Transfer to new parameters and the family certificate are assessed using the archived task instances: Table~\ref{tab:margin} applies each policy of Table~\ref{tab:records}, and the tied QAOA-1 alternative, to the other archived task instances. For HEA the selected policy retains all five law-affecting fields, so its zero residual elsewhere follows from the construction. For QAOA the choice between the two tied policies is decisive: $\{\mathrm{L,R,G,M}\}$, the set listed first in enumeration order, exceeds $\eps=0.05$ at four of thirteen instances, with maximum risk $0.0593$; $\{\mathrm{L,G,M,A}\}$ passes all thirteen with maximum risk $0.0444$, and the QAOA-2 policy passes all thirteen with maximum $0.0384$. The residual-risk rule was formulated after these failures were observed. Its numerical selection uses only design-instance data; comparisons with the further instances remain retrospective.

The finite-class formalism extends directly to a declared family of task instances. Let $\Omega_T$ be the disjoint union of the context classes of the tasks $t\in T$, with task identity a fixed, zero-cost, always-retained field. Then $\risk_{\Omega_T}(S)=\max_{t\in T}\risk_{\Omega_t}(S)$, and the finite-class arguments of Sections~\ref{sec:criterion} and~\ref{sec:separation} apply. Over the 28 archived QAOA instances, exhaustive enumeration of all 64 field sets gives $\{\mathrm{L,G,M,A}\}$ as the minimum-cost sufficient set at $\eps=0.05$, with family risk $0.0444$ attained at QAOA-1 seed 51. It is unique: the QAOA-2 design instance alone has singleton violations in each of L, G, M, and A, so every family-sufficient set contains all four. The HEA family of eight instances requires all five law-affecting fields. Supplementary Section~S3 summarizes the complete electronic results: family certificates, baseline sweeps, and both tie-break policies at every archived instance, including the four failures. The archive contains 11,664 contexts across 36 task instances with all laws, circuits, certificates, and witnesses.

\section{Finite-shot certification and scope}
\label{sec:scope}

The definition concerns underlying laws. A finite-shot comparison must additionally account for uncertainty in their estimates. Suppose estimates $\hat p_z$ satisfy simultaneous bounds $H(p_z,\hat p_z)\leq\eta_z$ with probability at least $1-\alpha$. The triangle inequality gives, for every pair at once,
\begin{equation}
 \begin{aligned}
 H(p_z,p_{z'})&\geq\max\{0,\hat H_{zz'}-\eta_z-\eta_{z'}\},\\
 H(p_z,p_{z'})&\leq\min\{1,\hat H_{zz'}+\eta_z+\eta_{z'}\},
 \end{aligned}
 \label{eq:confidence}
\end{equation}
where $\hat H_{zz'}=H(\hat p_z,\hat p_{z'})$. Setting diagonal terms to zero and taking maxima within record cells gives lower and upper risk bounds. A policy passes when its upper bound is at most $\eps$, fails when its lower bound exceeds $\eps$, and is unresolved otherwise. Inclusion-minimality additionally requires a lower bound above $\eps$ after deleting each retained field; unresolved deletions establish neither necessity nor minimality. Because the event is simultaneous, it also covers a policy selected using the same data.

The radius matters in practice. With $n_z$ independent shots at each of $M$ contexts on an alphabet of size $B$, Agrawal's bound on the moment generating function of the empirical relative entropy~\cite{Agrawal2020}, together with the inequality between relative entropy and Hellinger distance~\cite{vanErven2014}, gives, simultaneously with probability at least $1-\alpha$,
\begin{equation}
 \eta_z=\sqrt{1-\exp\!\left(-\frac{(B-1)\log2+\log(M/\alpha)}{n_z}\right)},
 \label{eq:radius}
\end{equation}
which decreases as $n_z^{-1/2}$ (Supplementary Section~S1.5). For $B=16$, $M=324$, and $\alpha=0.05$, this radius is $0.014$ at $10^5$ shots per context and $0.004$ at $10^6$; a radius of $0.025$ needs $30{,}669$ shots per context. This radius alone does not establish that a nontrivial policy passes: Eq.~\eqref{eq:confidence} adds two radii to each empirical pair distance. An elementary Hoeffding construction~\cite{Hoeffding1963} gives $0.250$ and $0.141$ at the same shot counts and needs $10^9$ shots for the same radius, because it passes through total variation and inherits an $n_z^{-1/4}$ rate. The shots that certify laws are distinct from the length $N$ of the records being compared in Section~\ref{sec:criterion}; the two enter the guarantee separately.

The relevant resource is the margin, not a radius in isolation. If $\widehat R$ is the empirical risk and $c=(B-1)\log2+\log(M/\alpha)$, a common shot count sufficient for the observed upper-bound test satisfies
\begin{equation}
 n\geq\frac{c}{-\log\!\left(1-[(\eps-\widehat R)/2]^2\right)},\qquad \widehat R<\eps.
 \label{eq:marginshots}
\end{equation}
For planning from a known model risk $R$, the stronger condition $4\eta\leq\eps-R$ ensures that the test passes on the simultaneous confidence event, because $\widehat R\leq R+2\eta$. At the QAOA-1 design instance this sufficient count is $4.42\times10^6$ shots per context for the residual-risk policy, versus $3.02\times10^9$ for the enumeration-order tie. These are conservative planning bounds for 324 contexts. The margin improves the sufficient acquisition budget by about a factor of 683. The analytic channel certificate, by contrast, requires no sampled output laws once its model and parameter ranges are justified.

\begin{table*}[!t]

\centering\small
\begin{tabular}{@{}lrrll@{}}\toprule
Comparison & Empirical risk & Simultaneous interval & Result & Cell sizes\\\midrule
Decoded, empty & 0.0272 & [0.0000, 0.0586] & pass & 6 \\
Raw, empty & 0.3873 & [0.3095, 0.4651] & fail & 6 \\
Raw, layout & 0.0333 & [0.0000, 0.1111] & pass & 3,3 \\
\bottomrule\end{tabular}
\caption{Fresh CT confirmation at the common tolerance $\eps=0.15$. Every passing
record shown pools multiple physical contexts; none is a singleton-cell pass.}
\label{tab:hardware}
\end{table*}

Distributional sufficiency also controls a declared bounded observable $f$ on the common outcome alphabet:
\begin{equation}
 \begin{split}
 |\mathbb{E}_{p_z}f-\mathbb{E}_{p_{z'}}f|
 &\leq (\max f-\min f)\,\dtv(p_z,p_{z'})\\
 &\leq \sqrt{2}(\max f-\min f)\,\eps
 \end{split}
 \label{eq:observable}
\end{equation}
for matching records. A direct observable criterion may be cheaper, as the study illustrates. Both criteria concern agreement between executions; accuracy relative to an ideal answer is a separate question, since two executions can agree and both be biased.

A finite certificate covers the declared contexts. Extending it to further devices or mechanisms requires evidence of the following kind. If every context in a larger class has an anchor in the certified class with the \emph{same retained record} and an outcome law within Hellinger distance $\eta$, then the larger-class risk is at most $\risk_\Omega(S)+2\eta$; Section~\ref{sec:precision} supplies model-based bounds for quantized continuous records. An anchor must have the same bin labels; rounding an exactly retained value generally fails that condition. Likewise, perturbing every law by at most $\eta$ changes every policy's risk by at most $2\eta$, so a policy with risk $\risk_\Omega(S)$ remains $\eps$-sufficient under perturbations up to $(\eps-\risk_\Omega(S))/2$, the allowance of Table~\ref{tab:margin}. Without such a coverage bound, testing additional contexts can falsify a proposed extension but cannot establish a universal one.

\section{A finite-context hardware confirmation}
\label{sec:hardware}

The hardware test asks whether an application decision requires the same
comparison record as the full bitstring law. On the six contexts tested, it does not: raw outputs need a two-label record, whereas for decoded outputs a single label suffices, with no additional context bits (Table~\ref{tab:hardware}). For this study, we used a six-variable conditional
boundary-refinement task derived from a binary computed-tomography (CT) formulation
\cite{kourousias_ct_2026}. Its input is a synthetic image; the measurements are real QPU outcomes. The exact reduction retains every
interaction with fixed exterior labels. One-layer QAOA prepares a fixed logical
task, and a decoder declared before acquisition retains two boundary variables,
$d(x)=(x_2,x_3)$. Raw and decoded alphabets therefore contain 64 and four outcomes.
The guarantee concerns these two boundary variables, not the entire segmentation.

A preceding scout selected this task for fresh confirmation. The objective,
Hamiltonian scale, angles, decoder and submitted native circuit bytes were kept
unchanged. Two six-qubit layouts on IBM Kingston~\cite{IBMProcessorTypes}, a Heron revision-2 processor,
were executed through IBM Quantum Platform in three new jobs on 29 September 2026.
Each of the six contexts
contains 32,768 application shots; controls add 3,072 shots. The three jobs used
60 provider-accounted QPU seconds. Scout shots do not enter these estimates.
Supplementary Section~S4 specifies the task, controls and archived execution record.
The preceding scout also ran the same frozen task on IQM Garnet~\cite{IQMGarnet} through IQM Resonance (Supplementary Section~S5); the reference label is the empirical mode in every Garnet context, but the decoded empty-record interval at $\eps=0.15$ remains unresolved, so no cross-provider pooling is certified.

The confirmation fixed $\eps=0.15$ for both alphabets and three joint targets:
decoded empty-record sufficiency, raw empty-record failure, and raw sufficiency
when layout is retained. The family allocation $\alpha=0.03$ is divided across
twelve law estimates, giving failure allocation $0.0025$ per law/alphabet.
Equation~\eqref{eq:radius}, with $\log(1/0.0025)$ in its logarithmic term, gives
raw and decoded radii 0.038915 and 0.015693. The common confidence
event covers both alphabets and all field subsets; their estimates need not be
independent of each other. The intervals remain conditional on independent,
stationary shots within each context. Fixed sample sizes, complete alphabets and
the absence of optional shot additions were specified before these measurements.

All three prespecified targets hold (Table~\ref{tab:hardware}). The decoded empty
record pools all six contexts, while the raw layout record pools three jobs in
each of two cells. Among subsets of backend, layout, compiler recipe and acquisition
block, with unit field costs, the sufficient minimum cost is consequently zero
for decoded laws and one for raw laws on this confidence event. The raw empty
record is the only cheaper policy and is excluded by a lower-bound witness;
backend and compiler are constant here.

The same evidence also determines the minimum label count over all deterministic
encodings of these six contexts. Raw empty-record failure excludes every one-label
encoding, while the passing layout partition supplies two labels. The decoded
empty record supplies one label. Thus exactly two labels are required for raw laws
and one for decoded laws: one versus zero additional fixed-length bits. Both counts hold on the same simultaneous confidence event. Hellinger contraction under decoding explains the direction of this change, but not the three-part conclusion at one common tolerance.

The two-pixel reference decision is the empirical mode in all six contexts. Basis-state controls show larger errors on the omitted variables in one layout, consistent with the raw/decoded contrast (Supplementary Section~S4). The tolerance $0.15$ is a distribution-comparison criterion, with a generic total-variation bound of about $0.211$. The result holds for these six contexts under the stated sampling model; validating the channel model of Section~\ref{sec:precision} on hardware is left to future work.

\section{Discussion and limitations}
\label{sec:discussion}

A sufficient record specifies a comparison at a declared statistical resolution. Finite classes identify necessary named fields through violating pairs. The channel certificate determines how much uncertainty may remain within continuous record cells, combining independent stochastic mechanisms with coherently accumulated rotations through affinity. A supplied channel inventory therefore gives a precision guarantee over a continuum of contexts and programmed parameters, with task identity fixed within each comparison.

The physical model is an input to that guarantee. Its branch maps, circuit structure and measurement must be common, and its rotation budget must cover the coherent differences. An unmodelled correlated environment, context-dependent measurement or compiler change can invalidate it. The conditional-fidelity form accommodates correlated classical flags when their joint affinity is controlled; the simpler product formula requires independence. Parameter-estimation uncertainty must be covered alongside quantization uncertainty. The attaining construction establishes sharpness for the general channel class, while the four-qubit examples quantify a sufficient envelope for their specified measurements.

Avoiding explicit register-dimension dependence does not remove dependence on the circuit inventory. Generating finite-class laws, enumerating pairs and optimizing field subsets remain separate computational costs. Error-mitigation quasi-probabilities require their own comparison object~\cite{Cai2023}. The hardware experiment establishes a finite-context retention distinction under its stated sampling model: a fixed application decoder changes the certified minimum label count at one common tolerance. Justifying a channel envelope independently, under measured drift, is the natural next experiment.

The finite-grid encoding example illustrates the cost of preserving field names: mechanism-specific labels can require more bits than a synthesized encoding of the distinctions visible in the output law. The appropriate representation depends on whether the objective is comparison, attribution or replay. Certified comparison records retain the declared tolerance and evidence class alongside the detailed execution archive needed to investigate how a result was produced.

In practice the method reduces to four steps: declare the task, the admissible contexts and the tolerance; select fields on finite classes by separating every tolerance-violating pair; set the bin widths of continuous fields from the channel certificate under the specified channel model; and archive the certificate with the record.

\begin{acknowledgments}
AI tools, including GPT-6 Astra and Fable 5.1, assisted mathematical consistency checks, acquisition and verification code, and manuscript drafting and editing. This included preparation and submission of jobs on IBM Quantum Platform and IQM Resonance under author-approved protocols. Exploratory certification prototypes informed the reasoning but are excluded from the reported results. Verification procedures are described in the Supplementary Information; the authors retain responsibility for the scientific content and its verification. This research received no specific funding.

\end{acknowledgments}

\section*{Data availability}
The data supporting this study are available in the Enodo archive~\cite{QuantumProvenanceDataset} (\href{https://doi.org/10.34965/enodo.i61153}{doi:10.34965/enodo.i61153}). It contains the complete finite-law study, analytic allocations, raw hardware outcomes for the synthetic CT task, native circuits and acquisition records. The Supplementary Information provides proofs and methods.

\section*{Code availability}
The same Enodo archive~\cite{QuantumProvenanceDataset} contains the offline reconstruction and verification code, acquisition-source snapshots, and instructions for installation and regeneration.

\section*{Competing interests}
The authors declare no competing interests.

\bibliography{references}

\begin{thebibliography}{33}%
\makeatletter
\providecommand \@ifxundefined [1]{%
 \@ifx{#1\undefined}
}%
\providecommand \@ifnum [1]{%
 \ifnum #1\expandafter \@firstoftwo
 \else \expandafter \@secondoftwo
 \fi
}%
\providecommand \@ifx [1]{%
 \ifx #1\expandafter \@firstoftwo
 \else \expandafter \@secondoftwo
 \fi
}%
\providecommand \natexlab [1]{#1}%
\providecommand \enquote  [1]{``#1''}%
\providecommand \bibnamefont  [1]{#1}%
\providecommand \bibfnamefont [1]{#1}%
\providecommand \citenamefont [1]{#1}%
\providecommand \href@noop [0]{\@secondoftwo}%
\providecommand \href [0]{\begingroup \@sanitize@url \@href}%
\providecommand \@href[1]{\@@startlink{#1}\@@href}%
\providecommand \@@href[1]{\endgroup#1\@@endlink}%
\providecommand \@sanitize@url [0]{\catcode `\\12\catcode `\$12\catcode
  `\&12\catcode `\#12\catcode `\^12\catcode `\_12\catcode `\%12\relax}%
\providecommand \@@startlink[1]{}%
\providecommand \@@endlink[0]{}%
\providecommand \url  [0]{\begingroup\@sanitize@url \@url }%
\providecommand \@url [1]{\endgroup\@href {#1}{\urlprefix }}%
\providecommand \urlprefix  [0]{URL }%
\providecommand \Eprint [0]{\href }%
\providecommand \doibase [0]{https://doi.org/}%
\providecommand \selectlanguage [0]{\@gobble}%
\providecommand \bibinfo  [0]{\@secondoftwo}%
\providecommand \bibfield  [0]{\@secondoftwo}%
\providecommand \translation [1]{[#1]}%
\providecommand \BibitemOpen [0]{}%
\providecommand \bibitemStop [0]{}%
\providecommand \bibitemNoStop [0]{.\EOS\space}%
\providecommand \EOS [0]{\spacefactor3000\relax}%
\providecommand \BibitemShut  [1]{\csname bibitem#1\endcsname}%
\let\auto@bib@innerbib\@empty
\bibitem [{\citenamefont {Peng}(2011)}]{Peng2011}%
  \BibitemOpen
  \bibfield  {author} {\bibinfo {author} {\bibfnamefont {R.~D.}\ \bibnamefont
  {Peng}},\ }\bibfield  {title} {\bibinfo {title} {Reproducible research in
  computational science},\ }\href {https://doi.org/10.1126/science.1213847}
  {\bibfield  {journal} {\bibinfo  {journal} {Science}\ }\textbf {\bibinfo
  {volume} {334}},\ \bibinfo {pages} {1226} (\bibinfo {year}
  {2011})}\BibitemShut {NoStop}%
\bibitem [{\citenamefont {K{\"o}ster}\ \emph {et~al.}(2026)\citenamefont
  {K{\"o}ster}, \citenamefont {Franz}, \citenamefont {Zec}, \citenamefont
  {Hoess}, \citenamefont {Ramsauer},\ and\ \citenamefont
  {Mauerer}}]{Koster2026}%
  \BibitemOpen
  \bibfield  {author} {\bibinfo {author} {\bibfnamefont {D.}~\bibnamefont
  {K{\"o}ster}}, \bibinfo {author} {\bibfnamefont {M.}~\bibnamefont {Franz}},
  \bibinfo {author} {\bibfnamefont {B.}~\bibnamefont {Zec}}, \bibinfo {author}
  {\bibfnamefont {N.}~\bibnamefont {Hoess}}, \bibinfo {author} {\bibfnamefont
  {R.}~\bibnamefont {Ramsauer}},\ and\ \bibinfo {author} {\bibfnamefont
  {W.}~\bibnamefont {Mauerer}},\ }\href
  {https://doi.org/10.48550/arXiv.2607.08348} {\bibinfo {title} {Works on my
  {QPU}: Reproducibility in quantum computing research}},\ \bibinfo
  {howpublished} {arXiv:2607.08348v2 [quant-ph]} (\bibinfo {year}
  {2026})\BibitemShut {NoStop}%
\bibitem [{\citenamefont {Dasgupta}\ and\ \citenamefont
  {Humble}(2022)}]{Dasgupta2022}%
  \BibitemOpen
  \bibfield  {author} {\bibinfo {author} {\bibfnamefont {S.}~\bibnamefont
  {Dasgupta}}\ and\ \bibinfo {author} {\bibfnamefont {T.~S.}\ \bibnamefont
  {Humble}},\ }\bibfield  {title} {\bibinfo {title} {Characterizing the
  reproducibility of noisy quantum circuits},\ }\href
  {https://doi.org/10.3390/e24020244} {\bibfield  {journal} {\bibinfo
  {journal} {Entropy}\ }\textbf {\bibinfo {volume} {24}},\ \bibinfo {pages}
  {244} (\bibinfo {year} {2022})}\BibitemShut {NoStop}%
\bibitem [{\citenamefont {Dasgupta}\ and\ \citenamefont
  {Humble}(2024)}]{Dasgupta2024}%
  \BibitemOpen
  \bibfield  {author} {\bibinfo {author} {\bibfnamefont {S.}~\bibnamefont
  {Dasgupta}}\ and\ \bibinfo {author} {\bibfnamefont {T.~S.}\ \bibnamefont
  {Humble}},\ }\bibfield  {title} {\bibinfo {title} {Impact of unreliable
  devices on stability of quantum computations},\ }\href
  {https://doi.org/10.1145/3682071} {\bibfield  {journal} {\bibinfo  {journal}
  {ACM Transactions on Quantum Computing}\ }\textbf {\bibinfo {volume} {5}},\
  \bibinfo {pages} {22:1} (\bibinfo {year} {2024})}\BibitemShut {NoStop}%
\bibitem [{\citenamefont {Rudinger}\ \emph {et~al.}(2019)\citenamefont
  {Rudinger}, \citenamefont {Proctor}, \citenamefont {Langharst}, \citenamefont
  {Sarovar}, \citenamefont {Young},\ and\ \citenamefont
  {Blume-Kohout}}]{Rudinger2019}%
  \BibitemOpen
  \bibfield  {author} {\bibinfo {author} {\bibfnamefont {K.}~\bibnamefont
  {Rudinger}}, \bibinfo {author} {\bibfnamefont {T.}~\bibnamefont {Proctor}},
  \bibinfo {author} {\bibfnamefont {D.}~\bibnamefont {Langharst}}, \bibinfo
  {author} {\bibfnamefont {M.}~\bibnamefont {Sarovar}}, \bibinfo {author}
  {\bibfnamefont {K.}~\bibnamefont {Young}},\ and\ \bibinfo {author}
  {\bibfnamefont {R.}~\bibnamefont {Blume-Kohout}},\ }\bibfield  {title}
  {\bibinfo {title} {Probing context-dependent errors in quantum processors},\
  }\href {https://doi.org/10.1103/PhysRevX.9.021045} {\bibfield  {journal}
  {\bibinfo  {journal} {Phys. Rev. X}\ }\textbf {\bibinfo {volume} {9}},\
  \bibinfo {pages} {021045} (\bibinfo {year} {2019})}\BibitemShut {NoStop}%
\bibitem [{\citenamefont {Mills}\ \emph {et~al.}(2026)\citenamefont {Mills},
  \citenamefont {Sireesh}, \citenamefont {Leichtle}, \citenamefont {Roffe},\
  and\ \citenamefont {Kashefi}}]{Mills2026}%
  \BibitemOpen
  \bibfield  {author} {\bibinfo {author} {\bibfnamefont {J.}~\bibnamefont
  {Mills}}, \bibinfo {author} {\bibfnamefont {A.}~\bibnamefont {Sireesh}},
  \bibinfo {author} {\bibfnamefont {D.}~\bibnamefont {Leichtle}}, \bibinfo
  {author} {\bibfnamefont {J.}~\bibnamefont {Roffe}},\ and\ \bibinfo {author}
  {\bibfnamefont {E.}~\bibnamefont {Kashefi}},\ }\bibfield  {title} {\bibinfo
  {title} {Logical accreditation: A framework for efficient certification of
  fault-tolerant computations},\ }\href {https://doi.org/10.1103/7pj4-gcbq}
  {\bibfield  {journal} {\bibinfo  {journal} {PRX Quantum}\ }\textbf {\bibinfo
  {volume} {7}},\ \bibinfo {pages} {033053} (\bibinfo {year}
  {2026})}\BibitemShut {NoStop}%
\bibitem [{\citenamefont {Moreau}\ and\ \citenamefont
  {Missier}(2013)}]{Moreau2013}%
  \BibitemOpen
  \bibfield  {author} {\bibinfo {author} {\bibfnamefont {L.}~\bibnamefont
  {Moreau}}\ and\ \bibinfo {author} {\bibfnamefont {P.}~\bibnamefont
  {Missier}},\ }\href {https://www.w3.org/TR/prov-dm/} {\bibinfo {title}
  {{PROV-DM}: The {PROV} data model}},\ \bibinfo {howpublished} {W3C
  Recommendation} (\bibinfo {year} {2013})\BibitemShut {NoStop}%
\bibitem [{\citenamefont {Soiland-Reyes}\ \emph {et~al.}(2022)\citenamefont
  {Soiland-Reyes}, \citenamefont {Sefton}, \citenamefont {Crosas} \emph
  {et~al.}}]{SoilandReyes2022}%
  \BibitemOpen
  \bibfield  {author} {\bibinfo {author} {\bibfnamefont {S.}~\bibnamefont
  {Soiland-Reyes}}, \bibinfo {author} {\bibfnamefont {P.}~\bibnamefont
  {Sefton}}, \bibinfo {author} {\bibfnamefont {M.}~\bibnamefont {Crosas}},
  \emph {et~al.},\ }\bibfield  {title} {\bibinfo {title} {Packaging research
  artefacts with {RO-Crate}},\ }\href {https://doi.org/10.3233/DS-210053}
  {\bibfield  {journal} {\bibinfo  {journal} {Data Science}\ }\textbf {\bibinfo
  {volume} {5}},\ \bibinfo {pages} {97} (\bibinfo {year} {2022})}\BibitemShut
  {NoStop}%
\bibitem [{\citenamefont {Weder}\ \emph {et~al.}(2021)\citenamefont {Weder},
  \citenamefont {Barzen}, \citenamefont {Leymann}, \citenamefont {Salm},\ and\
  \citenamefont {Wild}}]{Weder2021}%
  \BibitemOpen
  \bibfield  {author} {\bibinfo {author} {\bibfnamefont {B.}~\bibnamefont
  {Weder}}, \bibinfo {author} {\bibfnamefont {J.}~\bibnamefont {Barzen}},
  \bibinfo {author} {\bibfnamefont {F.}~\bibnamefont {Leymann}}, \bibinfo
  {author} {\bibfnamefont {M.}~\bibnamefont {Salm}},\ and\ \bibinfo {author}
  {\bibfnamefont {K.}~\bibnamefont {Wild}},\ }\bibfield  {title} {\bibinfo
  {title} {{QProv}: A provenance system for quantum computing},\ }\href
  {https://doi.org/10.1049/qtc2.12012} {\bibfield  {journal} {\bibinfo
  {journal} {IET Quantum Communication}\ }\textbf {\bibinfo {volume} {2}},\
  \bibinfo {pages} {171} (\bibinfo {year} {2021})}\BibitemShut {NoStop}%
\bibitem [{\citenamefont {Weder}\ \emph {et~al.}(2023)\citenamefont {Weder},
  \citenamefont {Barzen}, \citenamefont {Beisel},\ and\ \citenamefont
  {Leymann}}]{Weder2023}%
  \BibitemOpen
  \bibfield  {author} {\bibinfo {author} {\bibfnamefont {B.}~\bibnamefont
  {Weder}}, \bibinfo {author} {\bibfnamefont {J.}~\bibnamefont {Barzen}},
  \bibinfo {author} {\bibfnamefont {M.}~\bibnamefont {Beisel}},\ and\ \bibinfo
  {author} {\bibfnamefont {F.}~\bibnamefont {Leymann}},\ }\bibfield  {title}
  {\bibinfo {title} {Provenance-preserving analysis and rewrite of quantum
  workflows for hybrid quantum algorithms},\ }\href
  {https://doi.org/10.1007/s42979-022-01625-9} {\bibfield  {journal} {\bibinfo
  {journal} {SN Computer Science}\ }\textbf {\bibinfo {volume} {4}},\ \bibinfo
  {pages} {233} (\bibinfo {year} {2023})}\BibitemShut {NoStop}%
\bibitem [{\citenamefont {Huenchuleo}\ \emph {et~al.}(2026)\citenamefont
  {Huenchuleo}, \citenamefont {Sep{\'u}lveda},\ and\ \citenamefont
  {Fern{\'a}ndez}}]{Huenchuleo2026}%
  \BibitemOpen
  \bibfield  {author} {\bibinfo {author} {\bibfnamefont {N.}~\bibnamefont
  {Huenchuleo}}, \bibinfo {author} {\bibfnamefont {S.}~\bibnamefont
  {Sep{\'u}lveda}},\ and\ \bibinfo {author} {\bibfnamefont {A.}~\bibnamefont
  {Fern{\'a}ndez}},\ }\bibfield  {title} {\bibinfo {title} {{QC-MM}: A metadata
  and schema model for traceable quantum-circuit experiments},\ }\href
  {https://doi.org/10.3390/app16136346} {\bibfield  {journal} {\bibinfo
  {journal} {Applied Sciences}\ }\textbf {\bibinfo {volume} {16}},\ \bibinfo
  {pages} {6346} (\bibinfo {year} {2026})}\BibitemShut {NoStop}%
\bibitem [{\citenamefont {Stirbu}\ and\ \citenamefont {{Meijer van de
  Griend}}(2026)}]{Stirbu2026}%
  \BibitemOpen
  \bibfield  {author} {\bibinfo {author} {\bibfnamefont {V.}~\bibnamefont
  {Stirbu}}\ and\ \bibinfo {author} {\bibfnamefont {A.}~\bibnamefont {{Meijer
  van de Griend}}},\ }\href {https://doi.org/10.48550/arXiv.2608.05041}
  {\bibinfo {title} {Observing the quantum compiler through automatic
  experiment tracking for {Qiskit}}},\ \bibinfo {howpublished}
  {arXiv:2608.05041} (\bibinfo {year} {2026}),\ \bibinfo {note} {preprint,
  version 1, 5 August 2026}\BibitemShut {NoStop}%
\bibitem [{\citenamefont {Tishby}\ \emph {et~al.}(1999)\citenamefont {Tishby},
  \citenamefont {Pereira},\ and\ \citenamefont {Bialek}}]{Tishby1999}%
  \BibitemOpen
  \bibfield  {author} {\bibinfo {author} {\bibfnamefont {N.}~\bibnamefont
  {Tishby}}, \bibinfo {author} {\bibfnamefont {F.~C.}\ \bibnamefont
  {Pereira}},\ and\ \bibinfo {author} {\bibfnamefont {W.}~\bibnamefont
  {Bialek}},\ }\bibfield  {title} {\bibinfo {title} {The information bottleneck
  method},\ }in\ \href {https://arxiv.org/abs/physics/0004057} {\emph {\bibinfo
  {booktitle} {Proceedings of the 37th Annual Allerton Conference on
  Communication, Control, and Computing}}}\ (\bibinfo {year} {1999})\ pp.\
  \bibinfo {pages} {368--377}\BibitemShut {NoStop}%
\bibitem [{\citenamefont {Skowron}\ and\ \citenamefont
  {Rauszer}(1992)}]{Skowron1992}%
  \BibitemOpen
  \bibfield  {author} {\bibinfo {author} {\bibfnamefont {A.}~\bibnamefont
  {Skowron}}\ and\ \bibinfo {author} {\bibfnamefont {C.}~\bibnamefont
  {Rauszer}},\ }\bibfield  {title} {\bibinfo {title} {The discernibility
  matrices and functions in information systems},\ }in\ \href
  {https://doi.org/10.1007/978-94-015-7975-9_21} {\emph {\bibinfo {booktitle}
  {Intelligent Decision Support: Handbook of Applications and Advances of the
  Rough Sets Theory}}},\ \bibinfo {editor} {edited by\ \bibinfo {editor}
  {\bibfnamefont {R.}~\bibnamefont {S{\l}owi{\'n}ski}}}\ (\bibinfo  {publisher}
  {Kluwer Academic Publishers},\ \bibinfo {year} {1992})\ pp.\ \bibinfo {pages}
  {331--362}\BibitemShut {NoStop}%
\bibitem [{\citenamefont {Doshi}\ \emph {et~al.}(2010)\citenamefont {Doshi},
  \citenamefont {Shah}, \citenamefont {M{\'e}dard},\ and\ \citenamefont
  {Effros}}]{Doshi2010}%
  \BibitemOpen
  \bibfield  {author} {\bibinfo {author} {\bibfnamefont {V.}~\bibnamefont
  {Doshi}}, \bibinfo {author} {\bibfnamefont {D.}~\bibnamefont {Shah}},
  \bibinfo {author} {\bibfnamefont {M.}~\bibnamefont {M{\'e}dard}},\ and\
  \bibinfo {author} {\bibfnamefont {M.}~\bibnamefont {Effros}},\ }\bibfield
  {title} {\bibinfo {title} {Functional compression through graph coloring},\
  }\href {https://doi.org/10.1109/TIT.2010.2050835} {\bibfield  {journal}
  {\bibinfo  {journal} {IEEE Transactions on Information Theory}\ }\textbf
  {\bibinfo {volume} {56}},\ \bibinfo {pages} {3901} (\bibinfo {year}
  {2010})}\BibitemShut {NoStop}%
\bibitem [{\citenamefont {Braunstein}\ and\ \citenamefont
  {Caves}(1994)}]{Braunstein1994}%
  \BibitemOpen
  \bibfield  {author} {\bibinfo {author} {\bibfnamefont {S.~L.}\ \bibnamefont
  {Braunstein}}\ and\ \bibinfo {author} {\bibfnamefont {C.~M.}\ \bibnamefont
  {Caves}},\ }\bibfield  {title} {\bibinfo {title} {Statistical distance and
  the geometry of quantum states},\ }\href
  {https://doi.org/10.1103/PhysRevLett.72.3439} {\bibfield  {journal} {\bibinfo
   {journal} {Physical Review Letters}\ }\textbf {\bibinfo {volume} {72}},\
  \bibinfo {pages} {3439} (\bibinfo {year} {1994})}\BibitemShut {NoStop}%
\bibitem [{\citenamefont {Paris}(2009)}]{Paris2009}%
  \BibitemOpen
  \bibfield  {author} {\bibinfo {author} {\bibfnamefont {M.~G.~A.}\
  \bibnamefont {Paris}},\ }\bibfield  {title} {\bibinfo {title} {Quantum
  estimation for quantum technology},\ }\href
  {https://doi.org/10.1142/S0219749909004839} {\bibfield  {journal} {\bibinfo
  {journal} {International Journal of Quantum Information}\ }\textbf {\bibinfo
  {volume} {7}},\ \bibinfo {pages} {125} (\bibinfo {year} {2009})}\BibitemShut
  {NoStop}%
\bibitem [{\citenamefont {Kourousias}\ \emph
  {et~al.}(2026{\natexlab{a}})\citenamefont {Kourousias}, \citenamefont
  {Carrato}, \citenamefont {Pugliese}, \citenamefont {Hafner}, \citenamefont
  {Guzzi}, \citenamefont {Bill{\`e}}, \citenamefont {Bill{\`e}},\ and\
  \citenamefont {Dimitrakis}}]{QuantumProvenanceDataset}%
  \BibitemOpen
  \bibfield  {author} {\bibinfo {author} {\bibfnamefont {G.}~\bibnamefont
  {Kourousias}}, \bibinfo {author} {\bibfnamefont {S.}~\bibnamefont {Carrato}},
  \bibinfo {author} {\bibfnamefont {R.}~\bibnamefont {Pugliese}}, \bibinfo
  {author} {\bibfnamefont {A.}~\bibnamefont {Hafner}}, \bibinfo {author}
  {\bibfnamefont {F.}~\bibnamefont {Guzzi}}, \bibinfo {author} {\bibfnamefont
  {M.}~\bibnamefont {Bill{\`e}}}, \bibinfo {author} {\bibfnamefont
  {F.}~\bibnamefont {Bill{\`e}}},\ and\ \bibinfo {author} {\bibfnamefont
  {P.}~\bibnamefont {Dimitrakis}},\ }\href
  {https://doi.org/10.34965/enodo.i61153} {\bibinfo {title} {Dataset of
  sufficient quantum provenance}},\ \bibinfo {howpublished}
  {Enodo\DatasetDOIDisplay{10.34965/enodo.i61153}} (\bibinfo {year}
  {2026}{\natexlab{a}})\BibitemShut {NoStop}%
\bibitem [{\citenamefont {Watrous}(2018)}]{Watrous2018}%
  \BibitemOpen
  \bibfield  {author} {\bibinfo {author} {\bibfnamefont {J.}~\bibnamefont
  {Watrous}},\ }\href {https://doi.org/10.1017/9781316848142} {\emph {\bibinfo
  {title} {The Theory of Quantum Information}}}\ (\bibinfo  {publisher}
  {Cambridge University Press},\ \bibinfo {year} {2018})\BibitemShut {NoStop}%
\bibitem [{\citenamefont {Demkowicz-Dobrza{\'n}ski}\ \emph
  {et~al.}(2012)\citenamefont {Demkowicz-Dobrza{\'n}ski}, \citenamefont
  {Ko{\l}ody{\'n}ski},\ and\ \citenamefont {Gu{\c{t}}{\u{a}}}}]{Demkowicz2012}%
  \BibitemOpen
  \bibfield  {author} {\bibinfo {author} {\bibfnamefont {R.}~\bibnamefont
  {Demkowicz-Dobrza{\'n}ski}}, \bibinfo {author} {\bibfnamefont
  {J.}~\bibnamefont {Ko{\l}ody{\'n}ski}},\ and\ \bibinfo {author}
  {\bibfnamefont {M.}~\bibnamefont {Gu{\c{t}}{\u{a}}}},\ }\bibfield  {title}
  {\bibinfo {title} {The elusive {Heisenberg} limit in quantum-enhanced
  metrology},\ }\href {https://doi.org/10.1038/ncomms2067} {\bibfield
  {journal} {\bibinfo  {journal} {Nature Communications}\ }\textbf {\bibinfo
  {volume} {3}},\ \bibinfo {pages} {1063} (\bibinfo {year} {2012})}\BibitemShut
  {NoStop}%
\bibitem [{\citenamefont {Yuan}\ and\ \citenamefont {Fung}(2017)}]{Yuan2017}%
  \BibitemOpen
  \bibfield  {author} {\bibinfo {author} {\bibfnamefont {H.}~\bibnamefont
  {Yuan}}\ and\ \bibinfo {author} {\bibfnamefont {C.-H.~F.}\ \bibnamefont
  {Fung}},\ }\bibfield  {title} {\bibinfo {title} {Quantum parameter estimation
  with general dynamics},\ }\href {https://doi.org/10.1038/s41534-017-0014-6}
  {\bibfield  {journal} {\bibinfo  {journal} {npj Quantum Information}\
  }\textbf {\bibinfo {volume} {3}},\ \bibinfo {pages} {14} (\bibinfo {year}
  {2017})}\BibitemShut {NoStop}%
\bibitem [{\citenamefont {Tomita}\ and\ \citenamefont
  {Seki}(2003)}]{TomitaSeki2003}%
  \BibitemOpen
  \bibfield  {author} {\bibinfo {author} {\bibfnamefont {E.}~\bibnamefont
  {Tomita}}\ and\ \bibinfo {author} {\bibfnamefont {T.}~\bibnamefont {Seki}},\
  }\bibfield  {title} {\bibinfo {title} {An efficient branch-and-bound
  algorithm for finding a maximum clique},\ }in\ \href
  {https://doi.org/10.1007/3-540-45066-1_22} {\emph {\bibinfo {booktitle}
  {Discrete Mathematics and Theoretical Computer Science}}},\ \bibinfo {series}
  {Lecture Notes in Computer Science}, Vol.\ \bibinfo {volume} {2731}\
  (\bibinfo  {publisher} {Springer},\ \bibinfo {year} {2003})\ pp.\ \bibinfo
  {pages} {278--289}\BibitemShut {NoStop}%
\bibitem [{\citenamefont {Br{\'e}laz}(1979)}]{Brelaz1979}%
  \BibitemOpen
  \bibfield  {author} {\bibinfo {author} {\bibfnamefont {D.}~\bibnamefont
  {Br{\'e}laz}},\ }\bibfield  {title} {\bibinfo {title} {New methods to color
  the vertices of a graph},\ }\href {https://doi.org/10.1145/359094.359101}
  {\bibfield  {journal} {\bibinfo  {journal} {Communications of the ACM}\
  }\textbf {\bibinfo {volume} {22}},\ \bibinfo {pages} {251} (\bibinfo {year}
  {1979})}\BibitemShut {NoStop}%
\bibitem [{\citenamefont {Yang}\ \emph {et~al.}(2018)\citenamefont {Yang},
  \citenamefont {Bai}, \citenamefont {Chiribella},\ and\ \citenamefont
  {Hayashi}}]{Yang2018}%
  \BibitemOpen
  \bibfield  {author} {\bibinfo {author} {\bibfnamefont {Y.}~\bibnamefont
  {Yang}}, \bibinfo {author} {\bibfnamefont {G.}~\bibnamefont {Bai}}, \bibinfo
  {author} {\bibfnamefont {G.}~\bibnamefont {Chiribella}},\ and\ \bibinfo
  {author} {\bibfnamefont {M.}~\bibnamefont {Hayashi}},\ }\bibfield  {title}
  {\bibinfo {title} {Compression for quantum population coding},\ }\href
  {https://doi.org/10.1109/TIT.2017.2788407} {\bibfield  {journal} {\bibinfo
  {journal} {IEEE Transactions on Information Theory}\ }\textbf {\bibinfo
  {volume} {64}},\ \bibinfo {pages} {4766} (\bibinfo {year}
  {2018})}\BibitemShut {NoStop}%
\bibitem [{\citenamefont {Kandala}\ \emph {et~al.}(2017)\citenamefont
  {Kandala}, \citenamefont {Mezzacapo}, \citenamefont {Temme}, \citenamefont
  {Takita}, \citenamefont {Brink}, \citenamefont {Chow},\ and\ \citenamefont
  {Gambetta}}]{Kandala2017}%
  \BibitemOpen
  \bibfield  {author} {\bibinfo {author} {\bibfnamefont {A.}~\bibnamefont
  {Kandala}}, \bibinfo {author} {\bibfnamefont {A.}~\bibnamefont {Mezzacapo}},
  \bibinfo {author} {\bibfnamefont {K.}~\bibnamefont {Temme}}, \bibinfo
  {author} {\bibfnamefont {M.}~\bibnamefont {Takita}}, \bibinfo {author}
  {\bibfnamefont {M.}~\bibnamefont {Brink}}, \bibinfo {author} {\bibfnamefont
  {J.~M.}\ \bibnamefont {Chow}},\ and\ \bibinfo {author} {\bibfnamefont
  {J.~M.}\ \bibnamefont {Gambetta}},\ }\bibfield  {title} {\bibinfo {title}
  {Hardware-ef\-ficient variational quantum eigensolver for small molecules and
  quantum magnets},\ }\href {https://doi.org/10.1038/nature23879} {\bibfield
  {journal} {\bibinfo  {journal} {Nature}\ }\textbf {\bibinfo {volume} {549}},\
  \bibinfo {pages} {242} (\bibinfo {year} {2017})}\BibitemShut {NoStop}%
\bibitem [{\citenamefont {Farhi}\ \emph {et~al.}(2014)\citenamefont {Farhi},
  \citenamefont {Goldstone},\ and\ \citenamefont {Gutmann}}]{Farhi2014}%
  \BibitemOpen
  \bibfield  {author} {\bibinfo {author} {\bibfnamefont {E.}~\bibnamefont
  {Farhi}}, \bibinfo {author} {\bibfnamefont {J.}~\bibnamefont {Goldstone}},\
  and\ \bibinfo {author} {\bibfnamefont {S.}~\bibnamefont {Gutmann}},\ }\href
  {https://doi.org/10.48550/arXiv.1411.4028} {\bibinfo {title} {A quantum
  approximate optimization algorithm}},\ \bibinfo {howpublished}
  {arXiv:1411.4028} (\bibinfo {year} {2014})\BibitemShut {NoStop}%
\bibitem [{\citenamefont {Agrawal}(2020)}]{Agrawal2020}%
  \BibitemOpen
  \bibfield  {author} {\bibinfo {author} {\bibfnamefont {R.}~\bibnamefont
  {Agrawal}},\ }\bibfield  {title} {\bibinfo {title} {Finite-sample
  concentration of the multinomial in relative entropy},\ }\href
  {https://doi.org/10.1109/TIT.2020.2996134} {\bibfield  {journal} {\bibinfo
  {journal} {IEEE Transactions on Information Theory}\ }\textbf {\bibinfo
  {volume} {66}},\ \bibinfo {pages} {6297} (\bibinfo {year}
  {2020})}\BibitemShut {NoStop}%
\bibitem [{\citenamefont {van Erven}\ and\ \citenamefont
  {Harremo{\"e}s}(2014)}]{vanErven2014}%
  \BibitemOpen
  \bibfield  {author} {\bibinfo {author} {\bibfnamefont {T.}~\bibnamefont {van
  Erven}}\ and\ \bibinfo {author} {\bibfnamefont {P.}~\bibnamefont
  {Harremo{\"e}s}},\ }\bibfield  {title} {\bibinfo {title} {R{\'e}nyi
  divergence and {Kullback--Leibler} divergence},\ }\href
  {https://doi.org/10.1109/TIT.2014.2320500} {\bibfield  {journal} {\bibinfo
  {journal} {IEEE Transactions on Information Theory}\ }\textbf {\bibinfo
  {volume} {60}},\ \bibinfo {pages} {3797} (\bibinfo {year}
  {2014})}\BibitemShut {NoStop}%
\bibitem [{\citenamefont {Hoeffding}(1963)}]{Hoeffding1963}%
  \BibitemOpen
  \bibfield  {author} {\bibinfo {author} {\bibfnamefont {W.}~\bibnamefont
  {Hoeffding}},\ }\bibfield  {title} {\bibinfo {title} {Probability
  inequalities for sums of bounded random variables},\ }\href
  {https://doi.org/10.1080/01621459.1963.10500830} {\bibfield  {journal}
  {\bibinfo  {journal} {Journal of the American Statistical Association}\
  }\textbf {\bibinfo {volume} {58}},\ \bibinfo {pages} {13} (\bibinfo {year}
  {1963})}\BibitemShut {NoStop}%
\bibitem [{\citenamefont {Kourousias}\ \emph
  {et~al.}(2026{\natexlab{b}})\citenamefont {Kourousias}, \citenamefont
  {Carrato}, \citenamefont {Guzzi}, \citenamefont {Bill{\`e}}, \citenamefont
  {Hafner}, \citenamefont {Bill{\`e}}, \citenamefont {Pugliese},\ and\
  \citenamefont {Dimitrakis}}]{kourousias_ct_2026}%
  \BibitemOpen
  \bibfield  {author} {\bibinfo {author} {\bibfnamefont {G.}~\bibnamefont
  {Kourousias}}, \bibinfo {author} {\bibfnamefont {S.}~\bibnamefont {Carrato}},
  \bibinfo {author} {\bibfnamefont {F.}~\bibnamefont {Guzzi}}, \bibinfo
  {author} {\bibfnamefont {M.}~\bibnamefont {Bill{\`e}}}, \bibinfo {author}
  {\bibfnamefont {A.}~\bibnamefont {Hafner}}, \bibinfo {author} {\bibfnamefont
  {F.}~\bibnamefont {Bill{\`e}}}, \bibinfo {author} {\bibfnamefont
  {R.}~\bibnamefont {Pugliese}},\ and\ \bibinfo {author} {\bibfnamefont
  {P.}~\bibnamefont {Dimitrakis}},\ }\bibfield  {title} {\bibinfo {title}
  {{Quantum-native formulations for Computed Tomography: reconstruction,
  denoising and QUBO-based segmentation}},\ }\href
  {https://doi.org/10.1088/1748-0221/21/06/C06008} {\bibfield  {journal}
  {\bibinfo  {journal} {{Journal of Instrumentation}}\ }\textbf {\bibinfo
  {volume} {21}},\ \bibinfo {pages} {C06008} (\bibinfo {year}
  {2026}{\natexlab{b}})}\BibitemShut {NoStop}%
\bibitem [{\citenamefont {{IBM Quantum}}(2026)}]{IBMProcessorTypes}%
  \BibitemOpen
  \bibfield  {author} {\bibinfo {author} {\bibnamefont {{IBM Quantum}}},\
  }\href {https://quantum.cloud.ibm.com/docs/en/guides/processor-types}
  {\bibinfo {title} {Processor types}} (\bibinfo {year} {2026}),\ \bibinfo
  {note} {accessed 3 October 2026}\BibitemShut {NoStop}%
\bibitem [{\citenamefont {{IQM Quantum Computers}}(2026)}]{IQMGarnet}%
  \BibitemOpen
  \bibfield  {author} {\bibinfo {author} {\bibnamefont {{IQM Quantum
  Computers}}},\ }\href {https://www.iqmacademy.com/qpu/garnet/} {\bibinfo
  {title} {{Garnet}: 20-qubit quantum processor}} (\bibinfo {year} {2026}),\
  \bibinfo {note} {accessed 3 October 2026}\BibitemShut {NoStop}%
\bibitem [{\citenamefont {Cai}\ \emph {et~al.}(2023)\citenamefont {Cai},
  \citenamefont {Babbush}, \citenamefont {Benjamin}, \citenamefont {Endo},
  \citenamefont {Huggins}, \citenamefont {Li}, \citenamefont {McClean},\ and\
  \citenamefont {O'Brien}}]{Cai2023}%
  \BibitemOpen
  \bibfield  {author} {\bibinfo {author} {\bibfnamefont {Z.}~\bibnamefont
  {Cai}}, \bibinfo {author} {\bibfnamefont {R.}~\bibnamefont {Babbush}},
  \bibinfo {author} {\bibfnamefont {S.~C.}\ \bibnamefont {Benjamin}}, \bibinfo
  {author} {\bibfnamefont {S.}~\bibnamefont {Endo}}, \bibinfo {author}
  {\bibfnamefont {W.~J.}\ \bibnamefont {Huggins}}, \bibinfo {author}
  {\bibfnamefont {Y.}~\bibnamefont {Li}}, \bibinfo {author} {\bibfnamefont
  {J.~R.}\ \bibnamefont {McClean}},\ and\ \bibinfo {author} {\bibfnamefont
  {T.~E.}\ \bibnamefont {O'Brien}},\ }\bibfield  {title} {\bibinfo {title}
  {Quantum error mitigation},\ }\href
  {https://doi.org/10.1103/RevModPhys.95.045005} {\bibfield  {journal}
  {\bibinfo  {journal} {Reviews of Modern Physics}\ }\textbf {\bibinfo {volume}
  {95}},\ \bibinfo {pages} {045005} (\bibinfo {year} {2023})}\BibitemShut
  {NoStop}%
\end{thebibliography}%

\end{document}